# FULL DATA CONTROLLED WEB-BASED FEED AGGREGATOR

Haruna Isah

Department of Computing, School of Computing, Informatics and Media
University of Bradford, UK.
`latuji@computer.org`

## ABSTRACT

*Feed syndication is analogous to electronic newsletters, both are aimed at delivering feeds to subscribers; the difference is that while newsletter subscription requires e-mail and exposed you to spam and security challenges, feed syndication ensures that you only get what you requested for. This paper reports a review on the state of the art of feed aggregation technology and the development of a locally hosted web based feed aggregator as a research tool using the core features of WordPress; the software was further enhanced with plugins and widgets for dynamic content publishing, database and object caching, social web syndication, back-up and maintenance, among others. The results highlight the current developments in software re-use and describes; how open source content management systems can be used for both online and offline publishing, a means whereby feed aggregator users can control and share feed data, as well as how web developers can focus on extending the features of built-in software libraries in applications rather than reinventing the wheel.*

## 1. INTRODUCTION

Open source initiatives and the emergence of Web 2.0 have exposed the world to wide range of information sources. The World Wide Web has become packed with contents making it difficult for users to sort through and find what they are looking for. A means where by such information sources are bundled, filtered, sorted and delivered in a single useful and usable view is required. This led to the subject of syndication, usually between information sources (publishers) and destinations (subscribers). Web contents are separated from the presentation layer and passed as feeds to subscribers for aggregation into a single view (Aggregator). This technology is useful in various disciplines; from journalism to academic research, intelligence gathering, marketing and advertisement, communication within an organisation or professional groups as well as in media sharing. Feeds are designed for delivering web content updates, and are often referred to as RSS (Really Simple Syndication) which is a defined standard based on XML (eXtensible Markup Language) and have been released in several versions which includes; RSS 0.90, RSS 0.91, RSS 0.92, RSS 1.0, RSS 2.0 and Atom. Aggregators can be browser based, web based, desktop based or mobile based. Desktop and web based types are arguably the most preferred because of the simplicity of most of the interfaces as well as the user control capabilities. The web based and mobile types however possess another important feature, and that is everywhere access. This paper is aimed to review the current state of feed syndication as well as to develop a customised prototype web based feed aggregator with a database support for feed data control.

## 2. TECHNOLOGY REVIEW AND RELATED WORK

Some samples of widely used free web based aggregators were experimented and their features analysed and presented. Also a review on recent research work on feed aggregation was carried out as detailed below. This is a deliberate attempt to provide an insight on what had already been achieved on the topic in both the industry and the academia at the same time investigating what is still remaining as a challenge or research gap.

### 2.1. Related Work

Recent research effort on web based feed aggregators in the academia worth mentioning are the separate work put forward by [1] on the topic "A Study on Recommendation Features for an RSS Reader" as well as that of [2] on the topic "Performance Optimization of Web Based RSS Aggregator". The recorded successes of the former was their proposed four methods of ranking algorithm, a feature for recommending feeds to users; in a similar fashion, the success made by the latter on the challenges of handling and storing large amount of data was a recommendation (a prototype tested) for a large database with datasets to be linked to the aggregators. The database recommendation is one of the fundamental aims of this project. With the rapid growth of internet technologies [2, 17, 21] the social web is one other main feature lacking in most readers; this technology can aid research and information sharing when carefully integrated [1, 2, 20, 21, 22].

### 2.2. Technology Review

#### 2.2.1. Google Reader

Google reader is one of the popular web based feed aggregators which is free and available both offline (using Google gears) and online [3, 23]. It has the capability to read both RSS and atom feeds. Some of the Google reader's features that made it popular include; filtering, sorting, starring, sharing (now through Google+), OPML import and export, optional views (list and expanded), user settings and simple interface. Google Reader major drawbacks include:

- users don't have access to full control of feed data
- developers don't have access to source code

#### 2.2.2. Bloglines

Besides being free, Bloglines offers specialised widget support for sharing and publishing feeds (e.g. twitter, MySpace etc.). Bloglines have import and export, sharing, starring, sorting, filtering, iPhone version as well as application programming interface support. Detailed information can be found via [4]. Bloglines major drawbacks are similar to that of Google reader.

#### 2.2.3. MySyndicaat

A web based aggregator customised to support news professionals; as stated in [5], it has free and premium versions (for commercial users). The aggregator has both drawbacks mentioned in the previous sections. Its features however, include content aggregation, content editing, contentfiltering, content classification and re-classification, configuration management (including news radar management), republishing and multi-viewer support.

### 2.2.4. SuperFeedr

SuperFeedr is one of the aggregators that web developers and software professionals should appreciate. It has several API's; it also details the publisher-subscriber process of feed notification and update. It supports both RSS and Atom; notifications are received in JSON [6].Major drawback:

- users don't have access to full control of feed data
- although developers have more preference in SuperFeedr yet it is limited

### 2.2.5. rssLounge

rssLounge is an open source web based feed reader developed by Tobias Zeising. It is quite similar to Google reader yet having greater support for image blogging; rssLounge interface is Ajax-based and was built with Zend library. This aggregator can equally be hosted locally source codes are made free and available for download via [7].

## 3. SOFTWARE REUSE

Software reuse is the use of existing software products to develop or extend the features of another software system. Software requirement specifications, designs and implementation codes can be reused [8, 19].

### 3.1. Design Patterns

Design patterns are templates that developers use in communicating their knowledge and experience to others [8, 16]. The application of design patterns seems to be gaining popularity recently and its scope have widened becoming relevant in areas including software reuse, web development, problem solving, object technology learning, Human Computer Interaction, risk and project management etc. When carefully applied (otherwise performance will be greatly affected); design patterns can enhance software development processes by ensuring efforts are not duplicated and wheels are not re-invented. Details on Software Patterns are available via [8].

### 3.2. Observer

The observer or publish-subscribe design pattern is typical to feed syndication; it finds application in a program where one- to-many dependencies exist.  When the particular object (observable) that other objects depends on changes, all the objects (observers) that subscribed or depend on the observable are automatically notified [8].

## 4. CONTENT MANAGEMENT SYSTEM

Content Management System (CMS) in this context refers to server-side software designed to aid and simplify the creation and maintenance sites or any other information system. Popular types of CMS include:

- General-purpose/portals: e.g. Drupal, Joomla, TYPOS, MODX etc.
- Blogs: e.g. WordPress, Moveable Type, Expression Engine, Blogger etc.
- E-learning: e.g. Moodle, ATutor, Dot Learn etc.
- WIKIS: in conjunction with other tools e.g. Tiki Wiki, Media Wiki etc.

- Social Media: e.g. Dolphin, Rays, phpFox etc.

Sometimes these can be integrated into a very large complex one. WordPress, Joomla and Drupal have very large community support and usage [11, 24].

## 4.1. WordPress

WordPress is arguably one of the most popular open source blogging software with a broad array of options such as hosting, plugins, themes etc. Its simplicity made it easy to learn, extend, and customise.

### 4.1.1. Themes

Themes in WordPress refers to the presentation layer of an application, it contains style sheets and template files. Features common in most themes are widgetized areas, breadcrumb navigation, search engine capabilities. WordPress themes are not tied directly to the features and functionalities of the application hence switching between different themes may not affect the overall functionality of your site.

### 4.1.2. Plugins

Plugins are PHP scripts used to extend or change WordPress functionality; as such, its features and functionalities can be altered without affecting the core files. It therefore means "*plugins can extend WordPress to do almost anything you can imagine*" [10]. WordPress community is so matured that there is no shortage of plugins (free and premium) available via [11]. Plugins interact with WordPress application via Application Programming Interfaces (API). WordPress have two hooks (Actions and Filters) provided by the Plugin API that enable plugins to access specific features of the WordPress application [10, 11, 24].

The major advantages of using plugins may include:

- Ability for developers to modify its content without affecting other WordPress features
- It encourages code re-use and it is easy to update
- Easy to install, activate, configure, deactivate, uninstall etc.
- It has a huge community centred around its development

### 4.1.3. Widgets

These are special WordPress plugins that can be added, rearranged or removed by dragging and dropping from the theme sidebar through the administrative dashboard. Widgets provide an easy way to add more interactivity features to applications. Default widgets that often follow WordPress installation include: Text, calendar, links, archives, categories, recent comments, recent posts, custom menu, search, Meta, etc. [12].

## 5. REQUIREMENTS AND SPECIFICATIONS

### 5.1. Functional Requirements

The prototype feed aggregator should be able to:

- Display feeds from all the feed formats (RSS and Atom)
- Display the titles and summaries of articles in a feed with timestamp and other metadata.

- Display articles based on a special ordering (e.g. category, format, tags etc.)
- Authenticate user(s) and display feeds based on preferred settings
- Update feeds and database alterations automatically
- Automatically cache and minify database and objects
- Validate inputs and generate error messages where necessary

The software should be able to allow users to:

Login to the Administrator dashboardRead and share feeds
Register/add/subscribe new feedPublish and manage feeds and posts
Install and manage plugins and widgetsInstall and manage themes
Import or export feeds Create and manage pages and links
Moderate posts, pages, commentsRetire old articles
Manage database and perform CREATE, READ, UPDATE and DELETE functions

## 5.2. Non Functional Requirements

The non-functional requirements include; hardware and software constraints, quality issues, how useful and easy to use the system, to what extent are the system security and performance? Others include availability, robustness etc.

### 5.2.1. Software Requirement

Operating System: UNIX Systems, Windows
Browser: Firefox, Chrome, Internet Explorer, etc.
Applications: PHP 5.0 or higher, MySQL 5.0 or higher, Apache Server 2.0 or higher, FTPclient e.g. FileZillaWordPress 3.3.1 or higher and Dropbox

### 5.2.2. Hardware Requirement

Processor: 500 MHz, RAM: 512 MB, CACHE: 1 MB, Input Devices: Mouse & Keyboard, Output Device: Monitor with 60Hz frame rate.

### 5.2.3. Availability

Availability addresses the question of when and in what location is the system going to be available? The feed aggregator will be designed to be hosted on a local server in order to ensure a 24/7 day feed aggregation.

### 5.2.4. Flexibility

The system will be developed using WordPress technology where user files are stored separately and any customisation on such files may not affect its core files. An integrated application; popularly known as Cross platform Apache MySQL, PHP, and Perl (XAMPP) will be used at the course of the development process while local host apache server will be used to host the application.

### 5.2.5. Backup and Maintenance

WordPress provides an efficient way for users to back up their database, plugins, uploads, themes and site data and the data can be sent via the users e-mail, stored on the user's desktop or exported to Dropbox cloud account.

**5.2.6. Scalability**

The feed aggregator will be design with one of the world's most popular database, MySQL which is known to support extremely large data processing depending on the local hosing hardware capabilities etc.

**5.2.7. Usability**

How useful and useable the system is? How deep are the content pages as well as the general look and feel of the software? The feed aggregator will be design with an inherent simplicity in the user interface as well as the screen colours so that users can achieve what they desire with the system with ease and joy.

# 6. DESIGN

WordPress provides the mechanism for users to develop their themes and style sheets or chose any from hundreds available in the WordPress theme directory.A feed reader can be developed and customised with WordPress by extending its functionalities with built-in RSS parser or using RSS scrapers. The feeds will be directly stored and published in the wp_post table of the WordPress database.

## 6.1. Architecture

As an open source publishing platform, WordPress is a great tool for feed syndication [13]. The architecture of the aggregator will basically be dictated by the WordPress framework; the lean nature of the WordPress core files is such that developers can freely customise it to suit their needs.

## 6.2. Functional Components

Design issues to be addressed will include but not limited to the fundamental features of the application (syndication) to backup and maintenance as well as information sharing. The Plugins matching these features that were considered and the rational for the consideration are detailed below.

### 6.2.1. Syndication

This is the heart of the application, and it involves the use of the built-in RSS parser to fetch feeds from various sources to be rendered on a desired page. Feeds are stored as post in the wp_post table using aggregator plugins. Featured aggregator plugins available and being constantly updated for this function includes:

FeedWordPress**:** this plugin enables users to syndicate feeds into the MySQL database and publish them as posts. FeedWordPress is easy to use and can pull together contents from subscribed sources and make it available as a series of special posts to the public. The plugin is actively maintained and regularly updated.

WPeMatico:this plugin is used for auto-blogging and organises feeds into campaigns and automatically publish the feed as posts. The RSS feed fetching is carried out using the simple pie library included in WordPress while the core functions of WordPress is used for image processing. It stores feed into the WordPress wp_post database and it is actively maintained and regularly updated.

Syndicate Press:this plugin enables you to include feeds directly into WordPress posts, sidebars as widgets, pages or anywhere in the theme. Syndicate Press keeps you updated with latest news globally. The plugin is actively maintained and regularly updated.

Feedgrator:this widget enables you take a list of RSS feeds to be displayed as a single list. How the display should be is entirely up to the user. The widget is actively maintained and regularly updated.

WP RSS Aggregator:This plugin allows you import and merge multiple feeds using the built-in RSS parser (Simple Pie). Feeds are sorted by dates ordering from the latest to the oldest. The feed can be included within the theme or on posts and pages using the shortcode [wp-rss-aggregator]. Other aggregation plugins available in the WordPress library include: Flash Feed Scroll Reader, VikiSpot, Lifestream, Feedgeorge Wordpress Plugin, and Fantasy Sports Widget etc.

### 6.2.2. Caching

Caching plugins are desirable in web applications that queries database persistently; such process slows down the site thereby reducing performance. This is a fundamental requirement for this application since most plugins connects to the database.
*W3 Total Cache:*This plugin was designed to enhance the speed and user experience of WordPress sites. W3 Total cache reduces theme download time and provides a transparent Content Delivery Network (CDN) integration to WordPress sites. It is a great tool for minifying RSS feeds, posts and comments. Other caching plugins includes W3 Super Cache, DB Cache, and DB Cache Reloaded etc.

### 6.2.3. Backup and Maintenance

WordPress being actively maintained has a robust implicit security features and capabilities, however one among the numerous plugins that can be used for backup and maintenance is:
*WordPress Backup to Dropbox:* a must to have plugin that enables you to backup all your files on server periodically based on your scheduled settings. Its major strength is that your files will be stored in your Dropbox account and does not add more load to the server as other database backup's plugins such as WP-DB-Backup etc. does. It requires setting up a Dropbox account and linking the two.

### 6.2.4. Sharing

The significance of social web in information and content sharing, promotion, collection and storage cannot be underestimated. Facebook, Twitter, Flickr, MySpace etc. are great tools for content syndication [20].

*Social Sharing Toolkit:* An easy to configure plugin designed to aid content sharing via popular social networks such as Facebook, Twitter, and LinkedIn etc.

## 6.3. User Interface

The look and feel of the aggregator

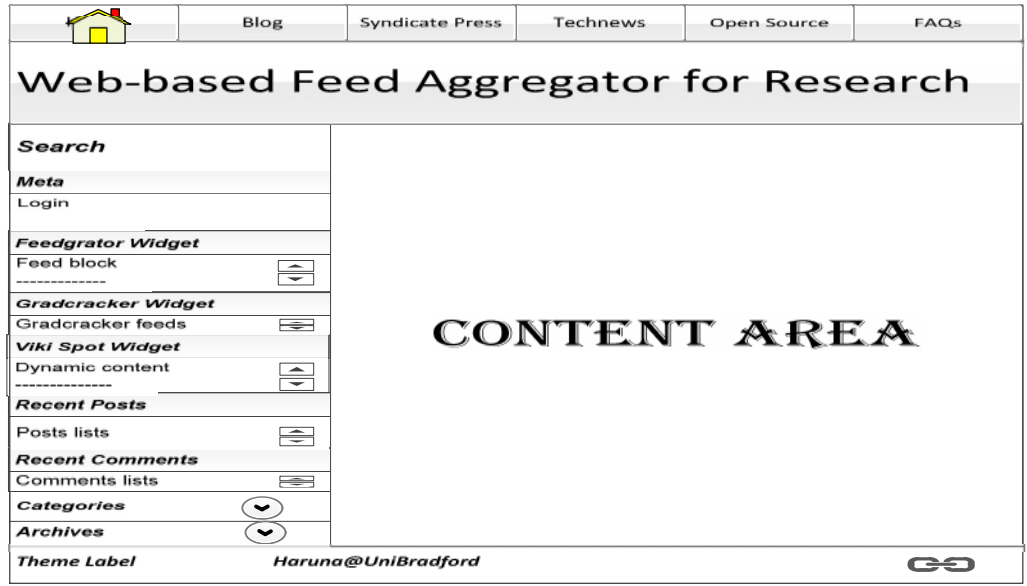

Figure 1.  User Interface Diagram

## 6.4. Database

The database for the feed aggregator is to be built up on the WordPress database which consists of eleven tables as adapted from the WordPress documentation. The tables include; wp_commentmeta, wp_comments, wp_links, wp_options, wp_postmeta, wp_posts, wp_terms, wp_term_relationships, wp_term_taxonomy, wp_usermeta, wp_users and their features are detailed in the WordPress.org codex section. Although all tables are useful yet the tables that are of fundamental significance for this application are the wp_posts and wp_postmeta where feeds and feeds data are store respectively as well as the wp_users and wp_usermeta tables that stores user details.

## 7. Implementation and Testing

The major aspects of the aggregator implementation includes; the simultaneous installation and configuration of the WordPress core files and database, this is followed by theme and plugins (widgets) installation and  configuration, then the creation of posts, pages, links, media, users etc. Settings and appearances are configured iteratively throughout the development process.

### 7.1. WordPress Core Files

The fundamental step in any software development is the file structure i.e. how folders and files are organised in servers. WordPress 3.4 default file structure like many other applications comprises of the; admin, content, and includes folders as well as the configuration, licensing, index etc. files in the same root folder.

### 7.1.1. Installed Theme

SimpleMarket version 1.1.1 was used for this application because it is minimal, elegant, HTML5 compliant and fit for any news related application. As described in its support forum; the theme has a single left side-bar with flexible-width and custom-background; it also supports single menu-bar, threaded-comments, and sticky-post.

### 7.1.2. Installed Plugins

The plugins that were installed includes: Feedgrator Test, FeedWordPress, WPeMatico, VikiSpot, and Syndicate Press (for feed syndication),others include; Akismet, WordPress Backup to Dropbox and W3 Total Cache for backup, caching and maintenance while Social Sharing Toolkit is meant to enhance collaboration and information sharing.

### 7.1.3. Generated Widgets

The following widgets were created upon the installation of the above plugins and are ready to be used in the theme widgetized areas: Akismet, Archives, Calendar, Categories, Custom Menu, Feedgrator, Links, List Category Posts, Meta, Pages, Recent Comments, Recent Posts, RSS, RSS feed, Search, Social Sharing Toolkit Follow Widget, Social Sharing Toolkit Share Widget, Tag Cloud, Text, VikiSpot (Content), VikiSpot (Stream).

## 7.2. Testing

Iterative plugins experimentation and black-box testing will be applied to test the most basic and crucial functions of the software, while an SQL interactive demonstrator will be used to perform database operations. The idea behind black box testing is that only a little knowledge about the application's data structure and algorithm is known, hence test inputs are checked on the application to verify the output [18]. Theactual output of a test case is compared with the expected output; and if equivalent, then that case is passed; else, the case has failed. The tables below describe the test cases developed to check the output of the fundamental functionalities of the web based feed aggregator.

Table 1.Syndication Test (WPeMatico)

| Step | Action | Expected Result |
|------|--------|-----------------|
| 1 | Click WPeMatico in the dashboard menu and complete campaigns>add new>activate>run now | The campaign ID, name, status, post type, count and run details displays in the dashboard and blog page |
| 2 | Click WPeMatico in the dashboard menu, select campaign to update and click any of copy, edit, delete, reset, deactivate or in bulk and click apply | The particular campaign selected should be updated as appropriate as well as feeds aggregated from same campaign |
| 3 | Click WPeMatico in the dashboard menu and complete settings, perform the necessary configuration and click save changes. | The settings made should be effective with immediate effect and but only future feeds can be affected |
| 4 | Click WPeMatico in the plugin list in the dashboard menu and complete any of deactivate or delete plugin | The Plugin and its contents should disappear |

Table 2.Syndication Test (feedgrator Widget)

| Step | Action | Expected Result |
|---|---|---|
| 1 | Click Appearances>Widget in the dashboard menu and drag the feedgrator widget to a desired position in the sidebar and then complete the settings with feed source etc.>save | Feeds displays in the widgetize area for valid sources based on the configuration |
| 2 | Click Appearances>Widget in the dashboard menu and select delete, close or drag the widget back to the non-widgetize area | The widget will disappear from the sidebar and the affected feeds will not show |
| 3 | Click Feedgrator Test in the plugin list in the dashboard and complete any of deactivate or delete plugin | The Plugin and its contents should disappear |

Table 3.Syndication Test (Syndicate Press)

| Step | Action | Expected Result |
|---|---|---|
| 1 | Click Settings>Syndicate Press in the dashboard menu and complete Output>enable-show content>Update Settings | Output aggregated feed content enabled and ready to publish any subscribed valid feed source |
| 2 | Click Settings>Syndicate Press in the dashboard menu and complete RSS Feeds> and List each RSS feed on a single line>Update Settings | Feeds should display on any post or page having the shortcode [sp# all] or accordance with user settings |
| 3 | Click Settings>Syndicate Press in the dashboard menu and complete Filters> inclusive and exclusive keyword filtering>Update Settings | The settings made should be effective affecting future feeds; the inclusive keyword filtering should be given precedence over exclusive keyword filtering when both are filled |
| 4 | Click Settings>Syndicate Press in the dashboard menu and complete Filters> input feed and formatted output caching>Update Settings | The appropriate caching configuration should become effective |
| 5 | Click Settings>Syndicate Press in the dashboard menu and complete Display Settings>Update Settings | The appropriate display configuration should become effective |
| 6 | Click Settings>Syndicate Press in the dashboard menu and complete Custom Formatting>Update Settings | The appropriate formatting configuration should become effective |
| 7 | Click Syndicate Press in the plugin list in the dashboard and complete any of deactivate or delete plugin | The Plugin and its contents should disappear |

Table 4.Syndication Test (FeedWordPress)

| Step | Action | Expected Result |
|------|--------|-----------------|
| 1 | Click Syndication in the dashboard menu and fill New source>add>update | Feed subscription should be successful after source validation and should show on the blog page |
| 2 | Click Syndication in the dashboard menu and complete Feed and Update Settings> save changes | The Feed, Updates, Update scheduling, Minimum interval, Time limit of updates, Updated posts and Advanced Settings should be effected as appropriate |
| 3 | Click Syndication in the dashboard menu and complete Syndicated Posts and Links Settings> save changes | The Posts, Syndicated Posts, Links, Formatting, Comments and Pings, Custom Post Settings and Types should be effected as appropriate |
| 4 | Click Syndication in the dashboard menu and fill Syndicated Author Settings> save changes | The Author and Syndicated Posts Settings should be effected as appropriate |
| 5 | Click Syndication in the dashboard menu and fill Categories and Tags Settings> save changes | The Categories, Feed Categories, Formats and Tags Settings should be effected as appropriate |
| 6 | Click Syndication in the dashboard menu and complete FeedWordPress Performance Settings> save changes | The FeedWordPress performance Settings should be updated as appropriate |
| 7 | Click Syndication in the dashboard menu and complete FeedWordPress Diagnostic Settings> save changes | The FeedWordPress Diagnostic Information, Display and Updates Settings should be updated as appropriate |

Table 5.Database related test case

| Step | Action | Expected Result |
|------|--------|-----------------|
| 1 | Visit phpMyAdmin page from a browser | The PhpMyAdmin login page displays |
| 2 | Enter valid database username and password | You should be directed to phpMyAdmin environment |
| 3 | Create, Read, Update, Delete (CRUD) and or import/ export etc. operation on any of the tables on the database | The operation performed should take effect immediately |

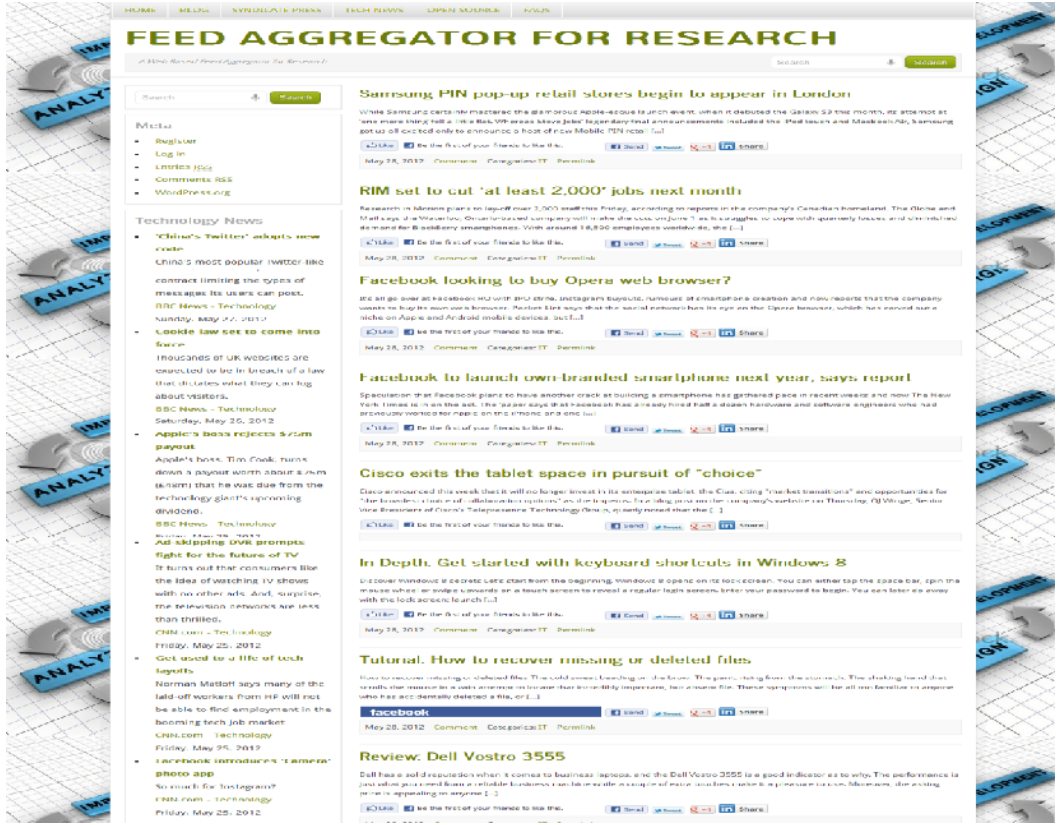

Figure 2.  Screen View of the Aggregator

## 8. RESULT ANALYSIS

Here the results of the test cases carried out are presented with the view of evaluating success and failures in the entire work. The test results for the most fundamental functionality of the software are presented with discussions in the tables below.

Table 6.  Result Analysis

| Test Case | Expected Result | Result Obtained | Remark |
|---|---|---|---|
| Syndication Test (WPeMatico) | Feed campaigns should be added to the database and be published as a blog post All other settings and configurations of the plugin should be performed | Subscribed feeds were pooled, added to database and published successfully Plugins functionality tested positive | Passed but no image support |
| Syndication Test (feedgrator Widget) | Feeds from subscribed sources should display based on user settings in the widgetize areas All other settings and configurations of the widget should be performed | Feeds displayed successfully in the sidebar and other settings and configurations of the widget should be performed | Passed but no image support |

| Syndication Test (Syndicate Press) | Feed should be published as a blog post or on any page having the plugin shortcode. All other settings and configurations of the plugin should be performed | Feed successfully published with a good image support in the syndicate press page However controlling count (feeds to be displayed per source) failed at the time of writing this report | Passed with good image support and feed count concerns |
|---|---|---|---|
| Syndication Test (FeedWordPress) | Feeds from subscribed sources should be added to the database and be published as a blog post All other settings and configurations of the plugin should be performed | Subscribed feeds were pooled, added to database and published successfully Plugins functionality tested positive | Passed but no image support |
| Admin Login | The Administrator should be able to login to admin dashboard and perform all admin related roles, error message should display on any attempt with invalid login details | All functionalities Successful, feeds were subscribed, profile edited, users, database, settings and updates operations performed | Passed |

## 9. CONCLUSIONS

Study of the various web-based feed aggregators available exposed two major gaps; the need for users to have full control of their feed data, the other being the need for the availability of the software source code and built-in library from developers point of view. This Web-basedfeed aggregator was designed majorly on open source tools (WordPress, Apache, PHP, and MySQL) allowing full feed data control and source code to users. The work is a major contribution in the field of software code reusability in that the entire software was implement with built-in themes, plugins and widgets. Feed Syndication was achieved using four plugins (FeedWordPress, WPeMatico, Syndicate Press, and Feedgrator Test) with the first two channeling feeds directly to the database (WordPress database and wp_post table), with Syndicate Press allowing the use of shortcode and filters to add feeds to pages or posts while Feedgrator test plugin creates a configurable feed display widget for sidebar applications. FeedWordPress plugin provides OPML (Outline Processor Markup Language) import of feed sources to simplify multiple subscriptions. CREATE, READ, UPDATE and DELETE operations were successfully tested using a simple SQL interactive demonstrator. Additional functionalities implemented include dynamic content publishing with VikiSpot, database and object caching with W3 Total Cache, social sharing to promote collaborative learning, backup and maintenance (update and backup with Dropbox). A fundamental requirement however is for users to have a basic knowledge of configuring, using and updating Apache, PHP, MySQL and WordPress.

Future work, specifically with regards to performance and further development of feed syndication plugins, themes and widgets is recommended. Features not successfully achieved include feed export, feed counts and loading performance.

## Author

Haruna Isah: Studied Electrical and Electronics Engineering between 2002 and 2008 at the University of Maiduguri, Nigeria. He is expected to graduate in July 2012 with MSc. in Software Engineering at the University of Bradford, United Kingdom. He has worked as a prototype developer with PEEMADI, a research institute of National Agency for Science and Engineering Infrastructure (NASENI) in Nigeria and currently working as a Graduate Assistant with the department of Electrical and Computer Engineering, Federal University of Technology Minna, Nigeria. He is a Professional member of ACM, IEEE and the Computer society of IEEE. His research interest includes Software Re-use, Data mining, Web Technologies, the Social Web and Artificial Intelligence.

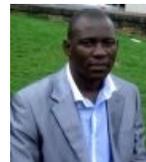